\documentclass[fleqn,usenatbib]{mnras}

\usepackage{newtxtext,newtxmath}

\usepackage[T1]{fontenc}

\DeclareRobustCommand{\VAN}[3]{#2}
\let\VANthebibliography\thebibliography
\def\thebibliography{\DeclareRobustCommand{\VAN}[3]{##3}\VANthebibliography}

\usepackage{graphicx}	
\usepackage{amsmath}	

\usepackage{threeparttable}

\title[TOI-2498\,b: A hot bloated super-Neptune within the Neptune desert]{TOI-2498\,b: A hot bloated super-Neptune within the Neptune desert}

\author[G Frame et al.]{\parbox{\textwidth}{\Large
Ginger~Frame$^{1,2}$\thanks{E-mail: ginger.frame@warwick.ac.uk},
David~J.~Armstrong$^{1,2}$,
Heather M.~Cegla$^{1,2}$,
Jorge~Fern\'andez~Fern\'andez$^{1,2}$,
Ares~Osborn$^{1,2}$,
Vardan Adibekyan$^{3,4}$,
Karen A.\ Collins$^{5}$,
Elisa Delgado Mena$^{3,4}$,
Steven Giacalone$^{6}$,
John F. Kielkopf$^{7}$,
Nuno C. Santos$^{3,4}$,
Sérgio G. Sousa$^{3,4}$,
Keivan G.\ Stassun$^{8}$,
Carl Ziegler$^{9}$,
David R. Anderson$^{1,2}$,
Susana C.C. Barros$^{3,4}$,
Daniel Bayliss$^{1,2}$
C\'{e}sar Brice\~{n}o$^{10}$,
Dennis M.\ Conti$^{11}$,
Courtney D. Dressing$^{6}$,
Xavier Dumusque$^{12}$,
Pedro~Figueira$^{3,4,12}$
William Fong$^{13}$,
Samuel Gill$^{1,2}$,
Faith Hawthorn$^{1,2}$,
Jon M. Jenkins$^{14}$
Eric L.\ N.\ Jensen$^{15}$,
Marcelo Aron F. Keniger$^{1,2}$,
David W.\ Latham$^{5}$,
Nicholas Law$^{16}$,
Jack J. Lissauer$^{17}$
Andrew W. Mann$^{16}$,
Louise D. Nielsen$^{18}$
Hugh Osborn$^{19}$,
Martin Paegert$^{5}$,
Sara Seager$^{13,20,21}$,
Richard P. Schwarz$^{5}$,
Avi Shporer$^{13}$,
Gregor Srdoc$^{22}$,
Paul A. Str{\o}m$^{1,2}$,
Joshua N. Winn$^{23}$,
Peter J. Wheatley$^{1,2}$
}
\vspace{0.2cm}
\\
$^{1}$Department of Physics, University of Warwick, Gibbet Hill Road, Coventry CV4 7AL, UK\\
$^{2}$Centre for Exoplanets and Habitability, University of Warwick, Gibbet Hill Road, Coventry CV4 7AL, UK\\
$^{3}$Instituto de Astrofísica e Ciências do Espaço, Universidade do Porto, CAUP, Rua das Estrelas, 4150-762 Porto, Portugal\\
$^{4}$Departamento de Física e Astronomia, Faculdade de Ciências, Universidade do Porto, Rua do Campo Alegre, 4169-007 Porto, Portugal\\
$^{5}$Center for Astrophysics \textbar \ Harvard \& Smithsonian, 60 Garden Street, Cambridge, MA 02138, USA\\
$^{6}$Department of Astronomy, University of California Berkeley, Berkeley, CA 94720, USA\\
$^{7}$Department of Physics and Astronomy, University of Louisville, Louisville, KY 40292, USA\\
$^{8}$Department of Physics and Astronomy, Vanderbilt University, Nashville, TN 37235, USA\\
$^{9}$Department of Physics, Engineering and Astronomy, Stephen F. Austin State University, 1936 North St, Nacogdoches, TX 75962, USA\\
$^{10}$Cerro Tololo Inter-American Observatory, Casilla 603, La Serena, Chile\\
$^{11}$American Association of Variable Star Observers, 185 Alewife Brook Parkway, Suite 410, Cambridge, MA 02138, USA\\
$^{12}$Observatoire Astronomique de l'Universit\'e de Gen\`eve, Chemin Pegasi 51b, CH-1290 Versoix, Switzerland\\
$^{13}$Department of Physics and Kavli Institute for Astrophysics and Space Research, Massachusetts Institute of Technology, Cambridge, MA 02139, USA\\
$^{14}$NASA Ames Research Center, Moffett Field CA 94035, USA\\
$^{15}$Department of Physics \& Astronomy, Swarthmore College, Swarthmore PA 19081, USA\\
$^{16}$Department of Physics and Astronomy, The University of North Carolina at Chapel Hill, Chapel Hill, NC 27599-3255, USA\\
$^{17}$Space Science \& Astrobiology Division, MS 245-3, NASA Ames Research Center, Moffett Field, CA 94035, USA\\
$^{18}$European Southern Observatory, Karl-Schwarzschild-Stra{\ss}e 2, 85748 Garching bei M{\"u}nchen, Germany\\
$^{19}$Physikalisches Institut, University of Bern, Gesellsschaftstrasse 6, 3012 Bern, Switzerland\\
$^{20}$Department of Earth, Atmospheric, and Planetary Sciences, Massachusetts Institute of Technology, Cambridge, MA 02139, USA\\
$^{21}$Department of Aeronautics and Astronautics, Massachusetts Institute of Technology, Cambridge, MA 02139, USA\\
$^{22}$Kotizarovci Observatory, Sarsoni 90, 51216 Viskovo, Croatia\\
$^{23}$Department of Astrophysical Sciences, Princeton University, Princeton, NJ 08544, USA\\
}

\date{Accepted 2023 May 10. Received 2023 May 5; in original form 2023 April 4}

\pubyear{2023}

\begin{document}
\label{firstpage}
\pagerange{\pageref{firstpage}--\pageref{lastpage}}
\maketitle
\begin{abstract}
We present the discovery and confirmation of a transiting hot, bloated Super-Neptune using photometry from {\it TESS} and LCOGT and radial velocity measurements from {\it HARPS}. The host star TOI-2498 is a V = 11.2, G-type (T\textsubscript{eff} = 5905 ± 12\,K) solar-like star with a mass of 1.12 ± 0.02 M\textsubscript{$\odot$} and a radius of 1.26 ± 0.04 R\textsubscript{$\odot$}. The planet, TOI-2498\,b, orbits the star with a period of 3.7 days, has a radius of 6.1 ± 0.3 R\textsubscript{$\oplus$}, and a mass of 35 ± 4 M\textsubscript{$\oplus$}. This results in a density of 0.86 ± 0.25\,g\,cm\textsuperscript{-3}. TOI-2498\,b resides on the edge of the Neptune desert; a region of mass-period parameter space in which there appears to be a dearth of planets. Therefore TOI-2498\,b is an interesting case to study to further understand the origins and boundaries of the Neptune desert. Through modelling the evaporation history, we determine that over its $\sim$3.6 Gyr lifespan, TOI-2498\,b has likely reduced from a Saturn sized planet to its current radius through photoevaporation. Moreover, TOI-2498\,b is a potential candidate for future atmospheric studies searching for species like water or sodium in the optical using high-resolution, and for carbon based molecules in the infra-red using JWST. 
\end{abstract}




\section{Introduction}

\noindent Since the seminal discovery of 51 Pegasi b \citep{mayor1995jupiter}, over 5000 exoplanets have been confirmed. Our search for exoplanets has reached a stage where detection capabilities are beginning to catch up to demands, meaning that we are able to look past detection biases and study the true exoplanet sample. This allows us to perform statistical studies on the exoplanet population, which forms a key role in gaining a better understanding of the formation of planetary systems. 

The {\it Kepler} \citep{borucki2010kepler} space mission was particularly successful in terms of the number of exoplanet detections, being responsible for over half of all confirmed exoplanets. {\it Kepler} was also successful in detecting smaller planets than were previously possible and found multitudes of super-Earths, filling in gaps in our knowledge of the exoplanet population. One key feature of the current exoplanet sample that was confirmed though {\it Kepler} discoveries is an apparent dearth of close orbit, Neptune sized planets \citep{szabo2011short, beauge2012emerging, helled2015possible, lundkvist2016hot, mazeh2016dearth, owen2018photoevaporation}. 

Continuing the work of the {\it Kepler} space telescope, the Transiting Exoplanet Survey Satellite ({\it TESS}) telescope \citep{sullivan2015transiting} was launched in 2018 and has since confirmed 285 exoplanets. Furthermore, {\it TESS} has been optimised to find planets around stars that are bright enough to be able to extract meaningful information from radial velocity observations. This enables us to be able to calculate mass and radius parameters of detected planets. Consequently, the {\it TESS} mission, along with spectroscopic instruments such as {\it HARPS} \citep{pepe2002harps}, have been integral in finding and characterising planets within the Neptune desert. In particular, the HARPS-NOMADS program endeavours to increase the number of planets in this category with precise mass and radius measurements. The objective of this pursuit is to be able to better study and understand the origins of the Neptune desert, and to better constrain the boundaries. Analysis of the boundaries of the Neptune desert could lead towards an improved understanding of formation and evolution of close orbit planets. 

This paper presents the detection of TOI-2498\,b, a hot super-Neptune transiting a G type star, residing on the edge of the Neptune desert. The observations used in this work include transits from {\it TESS} and LCO, spectroscopy from {\it HARPS} and spectral imaging from SOAR. The details of these observations are outlined in Section~\ref{obs}. Analysis of these observations, including joint modelling to constrain the planetary parameters and analysis to determine stellar parameters can be found in Section~\ref{analysis}. Section~\ref{results} contains the results of the analysis and a discussion involving the position of the planet in the Neptune desert and in mass-radius parameter space. This section also includes an analysis of the internal structure and evaporation history of the planet. Finally the conclusions of this work can be found in Section~\ref{conclusions}.

\section{Observations}
\label{obs}
TOI-2498 is a V = 11.2, G-type (T\textsubscript{eff} = 5905 ± 12\,K) solar-like star.  The properties of the star are set out in Table \ref{tab:stellar1}

\begin{table}
\renewcommand{\arraystretch}{1.2}

    \caption{Stellar parameters of TOI-2498}
    \label{tab:stellar1}
    \begin{tabular}{l|l|l}
    \hline 
    Stellar Parameters & Value & Source\\
    \hline

    {\bf Identifiers} & \\

    TIC ID & TIC-263179590 & TIC v8.2\\
    2MASS ID & J06213989+1115062 & 2MASS\\
    Gaia ID & 3330907293088717824 & \textit{Gaia} DR3\\

    \hline
    
    {\bf Astrometric Properties} & \\

    R.A. (deg) & $95.416220$ & \textit{Gaia} DR3\\
    DEC (deg) & $11.251645$  & \textit{Gaia} DR3\\
    Parallax (mas) & $3.61 \pm{ 0.05}$ & \textit{Gaia} DR3\\
    RV\textsubscript{sys} (km s$^{-1}$) & 8.36 & \textit{Gaia} DR3\\
    
    \hline
    {\bf Photometric Parameters} & \\
    B (mag) & $12.25 \pm 0.30$ & TIC v8.2\\
    V (mag) & $11.20 \pm 0.03$ & TIC v8.2\\
    G (mag) & $11.27 \pm 0.00$ & \textit{Gaia} DR3\\
    J (mag) & $10.28 \pm 0.03$ & TIC v8.2\\
    H (mag) & $10.00 \pm 0.02$ & TIC v8.2\\
    K (mag) & $9.93 \pm 0.02$ & TIC v8.2\\
    \textit{Gaia} BP (mag) & 11.61 & \textit{Gaia} DR3\\
    \textit{Gaia} RP (mag) & 10.80 & \textit{Gaia} DR3\\

    \hline
    
    \end{tabular}
    
    Sources: TIC v8.2 \citep{TIC}, 2MASS \citep{Skrutskie2006}, Gaia Data Release 3 \citep{GAIADR3}.
    
\end{table}

\subsection{TESS photometry}
\label{sec:TESS} 
TESS observed TOI-2498 for the first time during its primary mission (cycle 1) within sector 6, in Full Frame Images (FFIs) with a 30-min cadence. It was observed again during the first extended mission in sector 33 (cycle 3), with a 10 minute cadence FFI and a 2 minute cadence pixel stamp observation. 

The candidate was alerted as a TOI \citep[TESS Object of Interest,][]{guerrero2021tess} based on the identification of a transit signal from the SPOC pipeline \citep{jenkins2002impact, 10.1117/12.856764, jenkins2020kepler}. This pipeline uses the PDCSAP \citep[Presearch Data Conditioning Simple Aperture Photometry;][]{stumpe2012kepler, stumpe2014multiscale, smith2012kepler} light curves, which removes instrumental and some stellar trends from the SAP (Simple Aperture Photometry) data, but retains local features such as transits. The transit signal passed all the diagnostic tests conducted by Data Validation module \citep{Twicken:DVdiagnostics2018, Li:DVmodelFit2019}, and the difference image centroiding test located TOI-2498 to within 3.55 +/- 2.75 arcsec of the location of the transit source.

\begin{figure*}
\centering

\includegraphics[width=\textwidth, trim=0cm 1cm 0cm 1cm, clip] {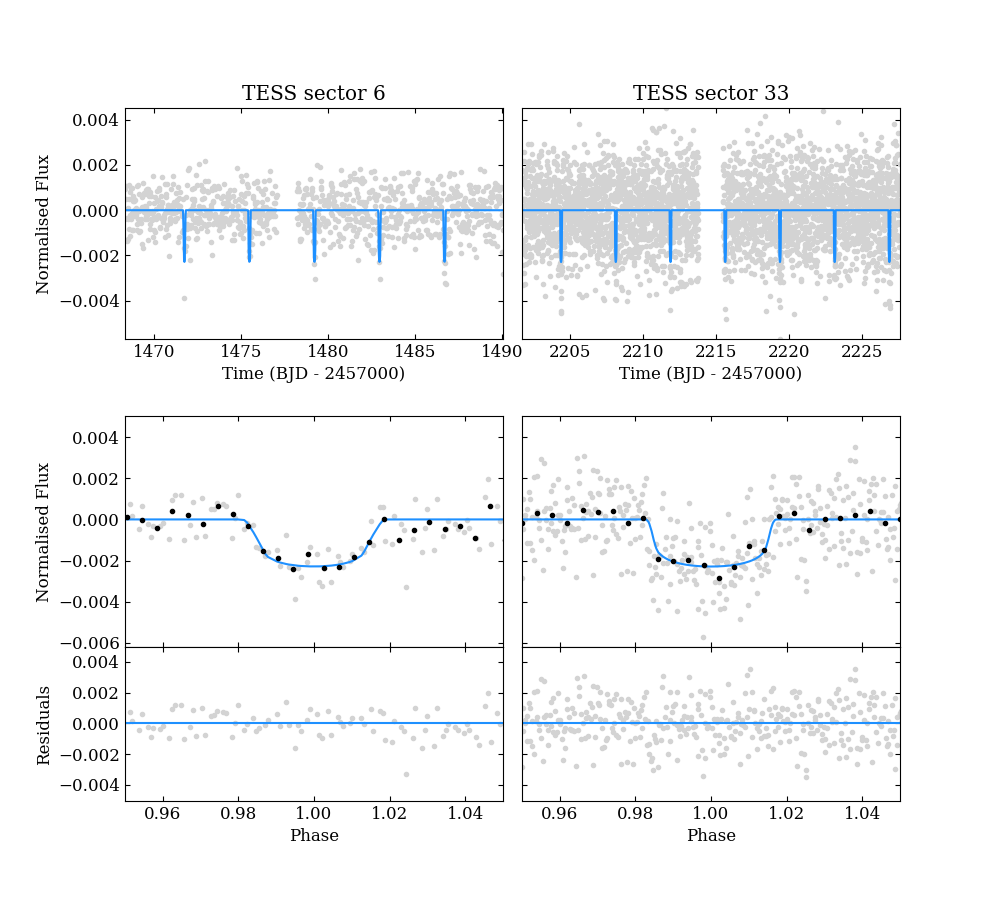}
\caption{The top plots show the {\it TESS} PDCSAP lightcurves from sector 6 (30 minute cadence) and sector 33 (2 minute cadence), flattened via a spline fit. Overplotted in blue is the best fit model resulting from the analysis described in section 3.2. The middle plots show this same data phase folded, with the binned data shown in black. Residuals of this fit can be found in the bottom plots. }

\label{fig:tess}
\end{figure*}

In this work, the PDCSAP lightcurves are flattened by removing the transits and fitting a spline to the data. The resulting flattened lightcurves are used in the joint model outlined in section \ref{subsection:model}. We present the normalised lightcurves along with the generated best-fit model in Figure \ref{fig:tess}. 

\begin{figure}
\centering

\includegraphics[width=\columnwidth]{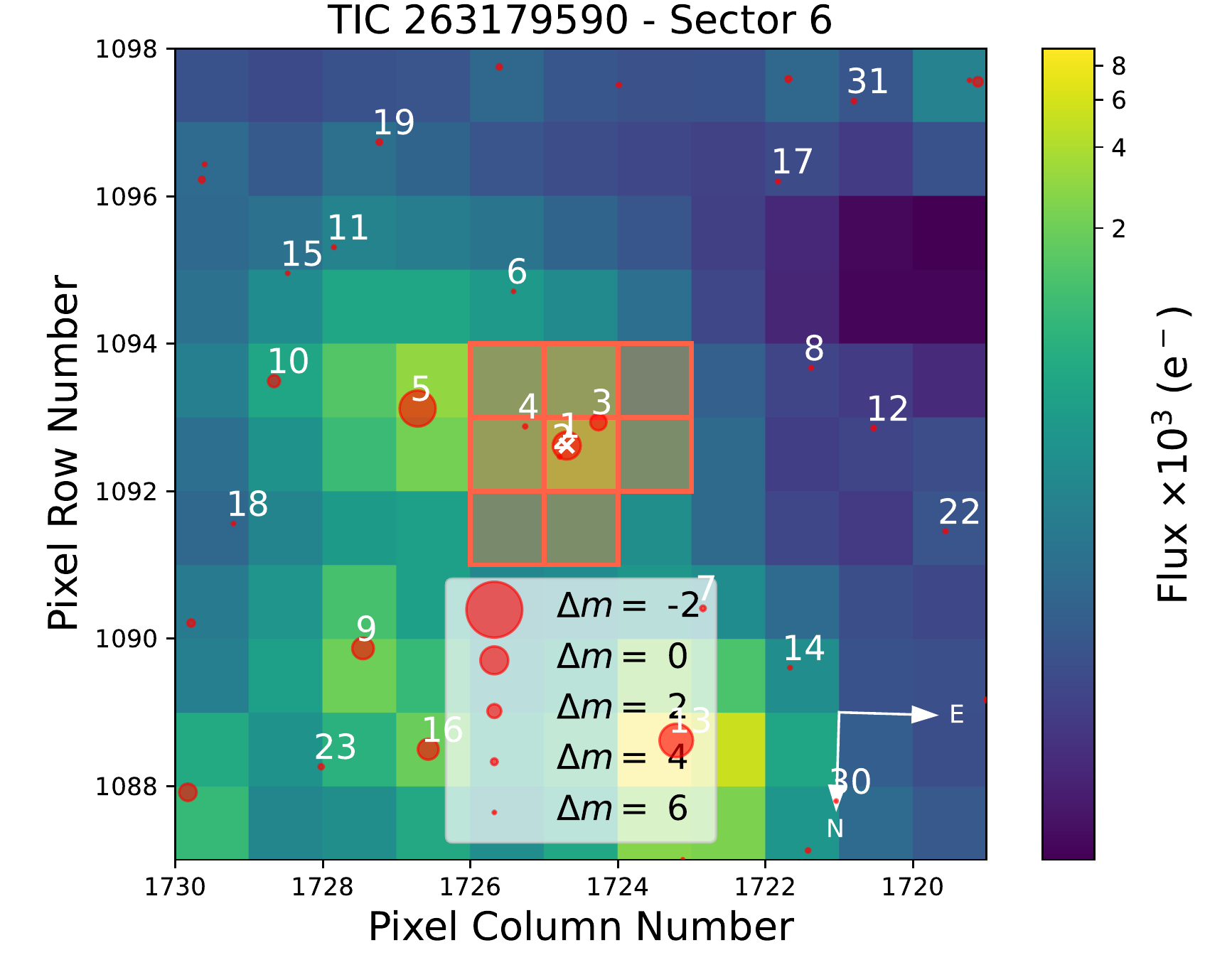}
\caption{Target Pixel File (TPF) from {\it TESS} sector 6. The shaded red squares indicate the aperture mask used by the SPOC pipeline. The target system (TOI-2498) is marked by a white cross, and the red circles mark nearby sources from the \textit{Gaia} DR3 catalogue, with sizes proportional to the difference in magnitude to the target.}
\label{fig:tpfgaia}
\end{figure}

Figure~\ref{fig:tpfgaia} shows the {\it TESS} Target Pixel File (TPF) created with \texttt{tpfplotter}\footnote{\url{https://github.com/jlillo/tpfplotter}} \citep{tpfplotter}, with TOI-2498 as the central object overlaid with nearby catalog sources. As can be seen in Figure \ref{fig:tpfgaia}, there are several sources of potential contamination around the target star, meaning it is likely that the {\it TESS} data is affected by blending of sources. To accurately measure the extent of this blending and correct for it, follow up ground based photometry was necessary; the details of these observations are described below.

\subsection{LCOGT follow-up photometry} 
\label{subsection:LCO}

The \textit{TESS} pixel scale is $\sim 21\arcsec$ pixel$^{-1}$ and photometric apertures typically extend out to roughly 1 arcminute, generally causing multiple stars to blend in the \textit{TESS} aperture. To determine the true source of the TOI-2498\,b detection in the \textit{TESS} data, improve the transit ephemeris, and check the SPOC pipeline transit depth after accounting for the crowding metric, we conducted ground-based lightcurve follow-up observations of the field around TOI-2498 as part of the {\textit TESS} Follow-up Observing Program\footnote{https://tess.mit.edu/followup} Sub Group 1 \citep[TFOP;][]{collins:2019}. We used the {\tt {\it TESS} Transit Finder}, which is a customized version of the {\tt Tapir} software package \citep{Jensen:2013}, to schedule our transit observations.

We observed a full predicted transit window of TOI-2498.01 on 2022 December 01 UT simultaneously using the Las Cumbres Observatory Global Telescope \citep[LCOGT;][]{Brown:2013} 1.0\,m network nodes at Teide Observatory and Cerro Tololo Inter-American Observatory. We also attempted a full transit observation on 2023 January 07 UT from the LCOGT 1.0\,m network node at Siding Spring Observatory, but deteriorating weather limited the in-transit coverage to about 60\% of the transit duration. The 1\,m telescopes are equipped with $4096\times4096$ SINISTRO cameras having an image scale of $0\farcs389$ per pixel, resulting in a $26\arcmin\times26\arcmin$ field of view. The images were calibrated by the standard LCOGT {\tt BANZAI} pipeline \citep{McCully:2018} and differential photometric data were extracted using {\tt AstroImageJ} \citep{Collins:2017}.

\begin{figure*}
\includegraphics[width=\textwidth] {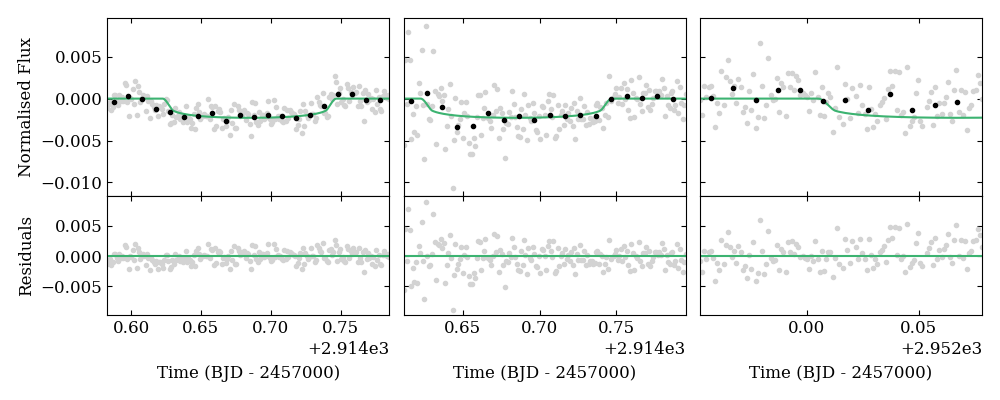}
\caption{LCOGT lightcurves from Teide Observatory (left), Cerro Tololo Inter-American Observatory (middle) and Siding Spring Observatory (right). The data is shown as grey points, with binned values in black. Overplotted is the best fit model from the joint fit described in section 3.2. The bottom plots show the residuals of this fit for each lightcurve. }

\label{fig:LCO}
\end{figure*}

We conducted focused observations to improve the separation of TOI-2498 from its nearest Gaia DR3 neighbor, which is $3\farcs8$ northwest and 5.7 magnitudes fainter in \textit{TESS} band. Typical stellar point spread functions (PSFs) in the images had full-width half-maximum (FWHM) in the range of 2-3 arcseconds. The Teide light curve had the smallest FWHM of $2\farcs16$ and the longest pre- and post-transit baseline of all three light curves, so resulted in the strongest constraints on transit model parameters. We first extracted differential light curves using circular photometric apertures with radius $5\farcs8$. In this case, the target star aperture included most of the flux of the nearest Gaia DR3 neighbor. A clear transit-like event was detected with a depth of $\sim 2000$\,ppm (deeper than predicted by the SPOC pipeline). We then extracted light curves using smaller $3\farcs1$ apertures that exclude the majority of the flux from the $3\farcs8$ neighbor. We again find a $\sim 2000$ ppm event in the smaller target aperture, confirming that the event occurs on-target relative to known Gaia DR3 stars. Since the flux contamination in the larger target star aperture is less than 1\%, we use the light curves from the the $5\farcs8$ apertures in the joint modeling in Section \ref{subsection:model}. We present the normalised lightcurves and generated best-fit model in Figure \ref{fig:LCO}.

\subsection{ULMT photometry}

We observed transit windows of TOI-2498 on 2021 February 26 and again on 2023 February 10 in the Sloan $r'$ band from the 0.61\,m University of Louisville Manner Telescope (ULMT) located at the Steward Observatory near Tucson, AZ. We used the {\tt {\it TESS} Transit Finder} to schedule these transit observations based on the ephemerides current at those times. The $4096\times4096$ SBIG STX-16803 camera on ULMT has an image scale of $0\farcs39$ per pixel, resulting in a $26\arcmin\times26\arcmin$ field of view. The images on both nights were slightly defocused and had typical stellar point-spread-functions with FWHM of $3\farcs4$ and $2\farcs9$ respectively. The images were calibrated and photometric data were extracted with {\tt AstroImageJ} \citep{Collins:2017} using circular aperture with radii $15\farcs9$ and $5\farcs9$ respectively.  In 2021, we tentatively detected a shallow event on-target  and did not find any obvious evidence of a blended nearby eclipsing binary that could be the cause of the {\it TESS} detection. In 2023, the shallow ingress was detected in detrended fitting. Due to the relatively high amount of noise compared the rest of our photometry, we decide not to include this data in our analysis.

\subsection{HARPS Radial Velocity observations}
\label{subsection:HARPS}

\noindent We acquired 16 spectra through the HARPS-NOMADS program (PI Armstrong, 1108.C-0697) using {\it HARPS} (the High Accuracy Radial velocity Planetary Searcher \cite{HARPS}). {\it HARPS} is an Echelle spectrograph mounted on the ESO 3.6-m telescope at the La Silla Observatory in Chile. The data were collected between 2022 September 30 to October 12 and 2022 November 15 to November 18. The {\it HARPS} High-Accuracy mode was utilised, with a fibre diameter of 1", a resolution of $R \approx 115,000$ and an exposure time of approximately 1800 seconds. This resulted in a signal to noise ratio (SNR) of between 30 and 35 per pixel at a wavelength of 550nm. The standard {\it HARPS} data reduction software outlined in \cite{lovis2007new} was used to reduce the raw {\it HARPS} data, in which a weighted cross correlation is applied with a G2 spectral mask. Measurements of the Radial Velocity (RV), FWHM, the line bisector span, the contrast of the Cross-Correalation Function (CCF) and the S-index are displayed in Table~\ref{tab:harpsobs}. The median photon noise per measurement for this dataset is 3.45 m\,s$^{-1}$. The last datapoint from the first set of observations is discarded due to anolamous results in RV error, FWHM, bisector span, contrast and S-index (see Table~\ref{tab:harpsobs}). This datapoint also has a lower SNR than the others, which we attribute to cloud coverage at the time of observation.

\begin{figure}
\includegraphics[width=\columnwidth, trim=0cm 1cm 0cm 1cm, clip] {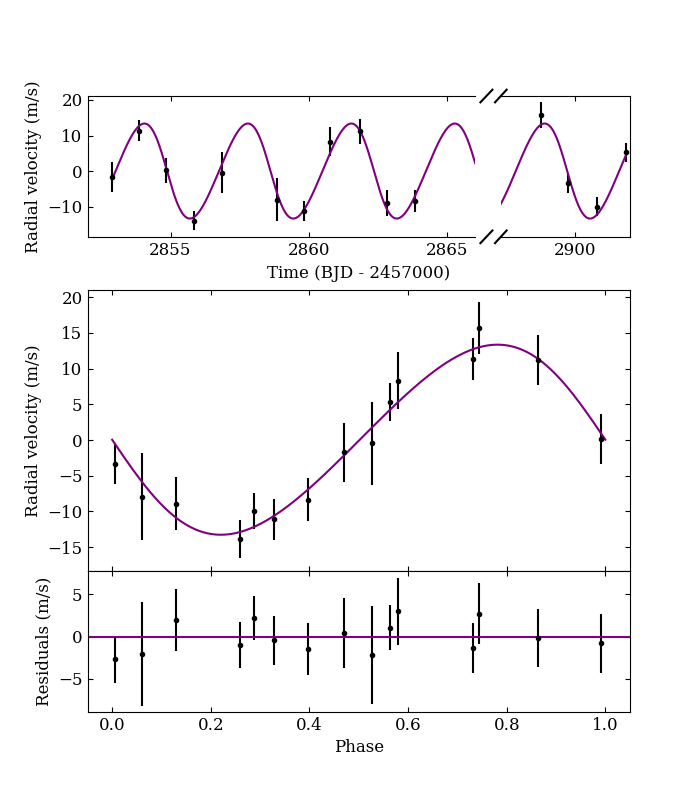}
\caption{Top: {\it HARPS} radial velocity data for TOI-2498. Overplotted is the best fit model obtained from the joint fit described in section 3.2, corresponding to the orbital period of TOI-2498\,b. Middle: The same data and model phase folded. Bottom: Residuals of the fit (observed RV - computed RV).}

\label{fig:HARPS}
\end{figure}

We present the {\it HARPS} radial velocity data along with the generated best-fit model (see Section~\ref{subsection:model} for details) in Figure \ref{fig:HARPS}.

\subsection{SOAR imaging}

\begin{figure}
\centering

\includegraphics[width=\columnwidth]{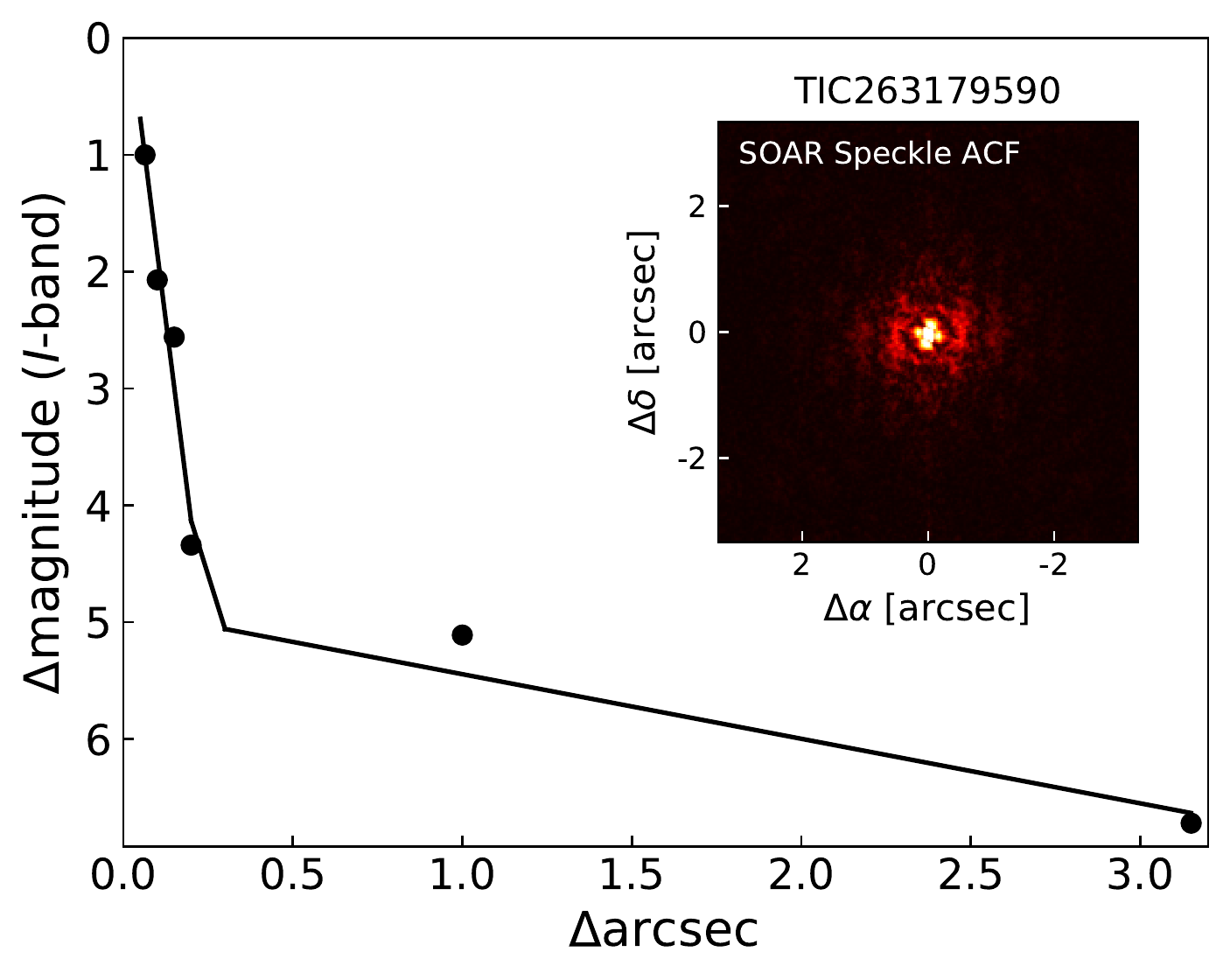}
\caption{Contrast curve computed from the high-resolution speckle observations in Cousins I-band on the 4.1 m Southern Astrophysical Research telescope. Inset is the speckle autocorrelation function centered on the target star. No bright companions are detected within 3" of TOI-2498 in this observation.}

\label{fig:SOAR}
\end{figure}

High-angular resolution imaging is needed to search for nearby sources that can contaminate the {\it TESS} photometry, resulting in an underestimated planetary radius, or be the source of astrophysical false positives, such as background eclipsing binaries. We searched for stellar companions to TOI-2498 with speckle imaging on the 4.1-m Southern Astrophysical Research (SOAR) telescope \citep{2018PASP..130c5002T} on 27 February 2021 UT, observing in Cousins I-band, a similar visible bandpass as TESS. This observation was sensitive to a 5.1-magnitude fainter star at an angular distance of 1 arcsec from the target. More details of the observations within the SOAR {\it TESS} survey are available in \cite{2020AJ....159...19Z}. The 5$\sigma$ detection sensitivity and speckle auto-correlation functions from the observations are shown in Figure \ref{fig:SOAR}. No nearby stars were detected within 3$\arcsec$ of TOI-2498 in the SOAR observations.

\subsection{Shane imaging}

\begin{figure}
\centering

\includegraphics[width=\columnwidth]{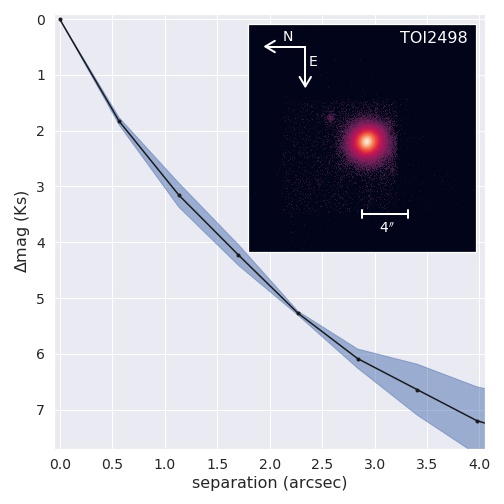}
\caption{Adaptive optics images of TIC 263179590 taken with the ShARCS camera on the Shane 3-meter telescope at Lick Observatory. We also present a contrast curve generated by calculating the median values (solid lines) and root-mean-square errors (blue, shaded regions) in annuli centered on each target, where the bin width of each annulus is equal to the full width at half max of the point spread function. TIC 715328024 ($\Delta$T = 5.7, sep = 3.8\arcsec) is visible to the N NW of the target star.}

\label{fig:Shane}
\end{figure}

We observed TOI-2498 on 2021 March 4 using the ShARCS camera on the Shane 3-meter telescope at Lick Observatory \citep{2012SPIE.8447E..3GK, 2014SPIE.9148E..05G, 2014SPIE.9148E..3AM}. Observations were taken with the Shane adaptive optics system in natural guide star mode to search for nearby, unresolved stellar companions. We collected a single sequence of observations using a $K_s$ filter ($\lambda_0 = 2.150$ $\mu$m, $\Delta \lambda = 0.320$ $\mu$m). We reduced the data using the publicly available \texttt{SImMER} pipeline \citep{2020AJ....160..287S, 2022PASP..134l4501S}.\footnote{https://github.com/arjunsavel/SImMER} Our reduced images and corresponding contrast curves are shown in Figure \ref{fig:Shane}. Our observations achieve a contrast of 2.9 at 1\arcsec and 4.8 at 2\arcsec. We detect the known star TIC 715328024 ($\Delta$T = 5.7, sep = 3.8\arcsec, PA = 328$^\circ$), but find no other nearby stellar companions within our detection limits.

\begin{center}
\begin{table*}
    \centering
    \caption{\textit{HARPS} spectroscopy. The row in \textbf{bold} text represents the data point that has been discarded due to anomalous results.}
    \label{tab:harpsobs}
    \begin{tabular}{l l l l l l l}  
    \hline
    \textbf{Time (BJD} & \textbf{RV} & \textbf{RV error} & \textbf{FWHM} & \textbf{Bisector span} & \textbf{Contrast} & \textbf{S-index\textsubscript{MW}} \\
    \textbf{-2457000)} & \textbf{(m/s)} & \textbf{(m/s)} & \textbf{(m/s)} & \textbf{(m/s)} &  & \\
    \hline
    2852.89471 & 7441.0 & 4.2 & 7262.1 & 17.8 & 46.57 & 0.13\\

    2853.86784 & 7454.1 & 2.9 & 7282.7 & -3.0 & 46.93 & 0.14 \\
    
    2854.83784 & 7443.0 & 3.5 & 7285.0 & 8.3 & 46.87 & 0.13 \\
    
    2855.84365 & 7429.0 & 2.7 & 7295.1 & 2.2 & 46.90 & 0.13 \\
    
    2856.84238 & 7442.3 & 5.8 & 7251.6 & 11.1 & 46.90 & 0.11 \\
    
    2858.83733 & 7434.8 & 6.1 & 7302.1 & 6.6 & 46.72 & 0.14 \\
    
    2859.83526 & 7431.7 & 2.9 & 7286.3 & -3.9 & 46.86 & 0.13 \\

    2860.77664 & 7451.0 & 4.0 & 7281.4 & -0.68 & 46.45 & 0.11 \\
    
    2861.83705 & 7453.9 & 3.4 & 7259.7 & 9.3 & 46.65 & 0.14 \\
    
    2862.83113 & 7433.9 & 3.7 & 7265.3 & 1.5 & 46.46 & 0.12 \\
    
    2863.83660 & 7434.4 & 3.0 & 7281.2 & 12.5 & 46.64 & 0.13 \\
    
    \textbf{2864.86315} & \textbf{7463.7} & \textbf{11.4} & \textbf{7176.2} & \textbf{40.1} & \textbf{44.29} & \textbf{0.06} \\
    
    2898.77390 & 7458.4 & 3.6 & 7285.9 & 8.1 & 46.77 & 0.12 \\
    
    2899.75712 & 7439.4 & 2.7 & 7276.7 & 1.1 & 46.96 & 0.14 \\
    
    2900.80942 & 7432.8 & 2.6 & 7287.5 & 11.6 & 46.81 & 0.12 \\
    
    2901.84129 & 7448.1 & 2.7 & 7289.8 & -2.3 & 46.86 & 0.13 \\
    
    \hline
    \end{tabular}
\end{table*}
\end{center}

\section{Analysis and Results}
\label{analysis}

\subsection{Stellar Analysis}

We used ARES+MOOG to derive the stellar atmospheric parameters T$_{\rm eff}$, $\log{g}$, microturbulence and [Fe/H]), and their respective uncertainties following the methodology described in \citet{Sousa-14, Santos-13}. We measured equivalent widths (EW) of iron lines on the combined {\it HARPS} spectrum of TOI-2498 using the ARES v2 code\footnote{The last version of ARES code (ARES v2) can be downloaded at http://www.astro.up.pt/$\sim$sousasag/ares} \citep{Sousa-15}. Then we used a minimization process where we assume ionization and excitation equilibrium to converge in the best set of spectroscopic parameters. This process makes use of a grid of Kurucz model atmospheres \citep{Kurucz-93} and the radiative transfer code MOOG \citep{Sneden-73}. 

Stellar abundances of the elements were derived using the classical curve-of-growth analysis method assuming local thermodynamic equilibrium and with the same codes and models that were used for the stellar parameters determinations. For the derivation of chemical abundances of refractory elements we closely followed the methods described in \cite{Adibekyan-12, Adibekyan-15, Delgado-17}. Abundances of the volatile elements, C and O, were derived following the method of \cite{Delgado-21, Bertrandelis-15}. We note that the oxygen lines are very weak and difficult to measure given the signal to noise ratio of the co-added spectrum and thus the final abundance has a large uncertainty. All the [X/H] ratios are obtained through a differential analysis with respect to a high S/N solar (Vesta) spectrum from HARPS. The stellar parameters and abundances of the elements are presented in Table~\ref{tab:stellar}. 

We find that the [X/Fe] ratios of TOI-2498 are typical for a thin disk star. Moreover, we used the chemical abundances of some elements to derive ages through the so-called chemical clocks (i.e. certain chemical abundance ratios which have a strong correlation for age). We applied the 3D formulas described in Table 10 of \citet{Delgado-19}, which also consider the variation in age produced by the effective temperature and iron abundance. The chemical clocks [Y/Mg], [Y/Zn], [Y/Ti], [Y/Si], [Y/Al], [Sr/Ti], [Sr/Mg] and [Sr/Si] were used from which we obtain a weighted average age of 3.6\,$\pm$\,1.1 Gyr. 

\subsubsection{Spectral Energy Distribution Analysis}

As an independent determination of the basic stellar parameters, we performed an analysis of the broadband spectral energy distribution (SED) of the star together with the {\it Gaia\/} DR3 parallax \citep[with no systematic offset applied; see, e.g.,][]{StassunTorres:2021}, in order to determine an empirical measurement of the stellar radius, following the procedures described in \citet{Stassun:2016,Stassun:2017,Stassun:2018}. We pulled the $B_T V_T$ magnitudes from {\it Tycho-2}, the $JHK_S$ magnitudes from {\it 2MASS}, the W1--W3 magnitudes from {\it WISE}, and the $G_{\rm BP} G_{\rm RP}$ magnitudes from {\it Gaia}. Together, the available photometry spans the full stellar SED over the wavelength range 0.4--10~$\mu$m.  

We performed a fit using PHOENIX stellar atmosphere models \citep{Husser:2013}, with the effective temperature ($T_{\rm eff}$), surface gravity ($\log g$), and metallicity ([Fe/H]) initially set to the spectroscopically determined values. The remaining free parameter is the extinction $A_V$, which we limited to maximum line-of-sight value from the Galactic dust maps of \citet{Schlegel:1998}. The resulting fit has a best-fit $A_V = 0.05 \pm 0.05$, with a reduced $\chi^2$ of 1.3. Integrating the (unreddened) model SED gives the bolometric flux at Earth, $F_{\rm bol} = 7.62 \pm 0.18 \times 10^{-10}$ erg~s$^{-1}$~cm$^{-2}$. Taking the $F_{\rm bol}$ together with the {\it Gaia\/} parallax, gives the bolometric luminosity, $L_{\rm bol} = 1.823 \pm 0.044$~L$_\odot$, which via the Stefan-Boltzmann relation gives the stellar radius, $R_\star = 1.291 \pm 0.018$~R$_\odot$. In addition, we can estimate the stellar mass from the empirical eclipsing-binary relations of \citet{Torres:2010}, giving $M_\star = 1.17 \pm 0.07$~M$_\odot$. 

Finally, we performed a fit allowing the stellar parameters to vary, finding only the metallicity differed marginally from the spectroscopically determined value, [Fe/H] = $0.05 \pm 0.15$, with an improved reduced $\chi^2$ of 0.8. In that case, the resulting stellar properties are as follows: $A_V = 0.05 \pm 0.05$, $F_{\rm bol} = 7.77 \pm 0.18 \times 10^{-10}$ erg~s$^{-1}$~cm$^{-2}$, $L_{\rm bol} = 1.860 \pm 0.045$~L$_\odot$, $R_\star = 1.305 \pm 0.018$~R$_\odot$, and $M_\star = 1.15 \pm 0.07$~M$_\odot$. 

As all determined stellar parameters were in agreement, we adopt the ARES+MOOG parameter set for consistency.

\subsection{Joint modelling}
\label{subsection:model}
The photometry from both {\it TESS} and LCO and spectroscopy from HARPS, described in Section~\ref{obs}, were combined in a joint fit using the \texttt{exoplanet} \citep{exoplanet:exoplanet} code framework. This package also makes use of \texttt{starry} \citep{luger2019starry} and \texttt{PYMC3} \citep{salvatier2016pymc3}. 

\subsubsection{Photometry}
\label{subsection:photometry}

We include the flattened photometric data from the two {\it TESS} sectors described in section \ref{sec:TESS} and the two full and one partial transit from LCOGT described in section \ref{subsection:LCO}. 

Firstly, we normalise all photometry by dividing each lightcurve by the median of the out-of-transit flux and subtracting one. Each transit is then modelled through \texttt{exoplanet} as a Keplerian orbit, with quadratic limb-darkening parameterisation from \cite{exoplanet:kipping13}. 
The Keplerian orbit model is parameterised in terms of stellar mass ($M_{\star}$) and radius ($R_{\star}$) in solar units, the orbital period P in days, central transit ephemeris T\textsubscript{C} in TBJD (BJD - 2457000), impact parameter $b$, eccentricity $e$ and argument of periastron $\omega$ in radians. Transit models are generated using \texttt{starry}, and include planetary radius ($R_{p}/R_{\star})$ and the exposure time for the relevant instrument. 

\subsubsection{Spectroscopy}

We include the spectroscopic data from {\it HARPS} described in section \ref{subsection:HARPS}. We find no need to fit a GP model as periodogram analysis through the DACE\footnote{DACE can be accessed at https://dace.unige.ch} platform shows no indication of stellar activity or signals from stellar rotation, suggesting that TOI-2498 is magnetically quiet. We fit for the semi-amplitude K of the RV signal, starting with a widely set uniform prior centered around the value found via analysis using DACE (13.37 m s$^{-1}$). We also fit for {\it HARPS} instrument offset, and implement a noise model which adds a white noise (jitter) term in quadrature with the nominal RV uncertainties. 

\subsubsection{Joint fit results}

We first maximise the log probability of the \texttt{PYMC3} model, and use the resulting parameters as our initial fit values. We then use a No U-Turn Sampler (NUTS) variant of the Hamiltonian Monte Carlo (HMC) algorithm to draw samples from the posterior chain, for which we use 20 chains with 20,000 steps. We treat the first 4000 samples drawn as burn-in and subsequently discard them. 

Through the results of this analysis, we determine that TOI-2498\,b is a Super-Neptune of radius 6.06 ± 0.29 R\textsubscript{$\oplus$} and mass 34.62 ± 4.10 M\textsubscript{$\oplus$}. From this we infer a bulk density of 0.86 ± 0.25 gcm\textsuperscript{-3}. We present all parameter best-fit results in Table \ref{tab:planet}.

\begin{table}
\renewcommand{\arraystretch}{1.3}

    \caption{Stellar parameters of TOI-2498 derived from this work}
    \label{tab:stellar}
    \begin{tabular}{l|l}
    \hline 
    Stellar Parameters & Value\\
    \hline
    
    Mass (M$_{\odot}$) & $1.12 \pm 0.02$\\
    Radius (R$_{\odot}$) & $1.26 \pm 0.04$\\
    Density (g cm\textsuperscript{-3}) & $0.79^{+0.10}_{-0.08}$\\
    log g & $4.297 \pm 0.021$ \\
    T$_{\rm eff}$ (K) & $5905 \pm 12$ \\
    v\textsubscript{turb} (km s\textsuperscript{-1}) & $1.125 \pm 0.016$ \\
    Age (Gyr) & $3.6 \pm 1.1$ \\

    \hline
    {\bf Stellar abundances} & \\

    [Fe/H] (dex) & $0.167^{+0.010}_{-0.041}$  \\
    
    [C/H] (dex) & $0.071 \pm 0.029$  \\
    
    [O/H] (dex) & $0.028 \pm 0.160$  \\
    
    [Mg/H] (dex) & $0.19 \pm 0.06$  \\
    
    [Al/H] (dex) & $0.22 \pm 0.07$  \\
    
    [Si/H] (dex) & $0.16 \pm 0.03$   \\
    
    [Ti/H] (dex) & $0.18 \pm 0.03$  \\
    
    [Ni/H] (dex) & $0.17 \pm 0.02$  \\
    
    [Cu/H] (dex) & $0.213 \pm 0.025$  \\
    
    [Zn/H] (dex) & $0.144 \pm 0.016$  \\
    
    [Sr/H] (dex) & $0.146 \pm 0.077$ \\
    
    [Y/H] (dex) & $0.171 \pm 0.056$  \\
    
    [Zr/H] (dex) & $0.089 \pm 0.065$  \\
    
    [Ba/H] (dex) & $0.29 \pm 0.028$\\
    
    [Ce/H] (dex) & $0.137 \pm 0.059$  \\
    
    [Nd/H] (dex) & $0.135 \pm 0.033$  \\

    \hline
    
    \end{tabular}
    
\end{table}

\begin{table}

\caption{Prior distributions of fitted parameters}
    
    \begin{tabular}{lll}
      \hline
      {\bf Parameter} & {\bf Prior distribution} & {\bf Fitted value}\\
      \hline
      TESS Sector 6 mean & $\mathcal{N}$(0, 1) & $0.00005 \pm 0.00002$ \\
      TESS Sector 33 mean & $\mathcal{N}$(0, 1) & $0.00002 \pm 0.00002$\\
      Contamination $^{*}$ & $\mathcal{N}$(0.66, 0.2) & $0.63 \pm 0.05$\\
      \hline
      LCO Teide mean & $\mathcal{N}$(0, 1) & $-0.0001 \pm 0.00009$\\
      LCO CTIO mean & $\mathcal{N}$(0, 1) & $-0.0003 \pm 0.0001$\\
      LCO SSO mean & $\mathcal{N}$(0, 1) & $0.0004 \pm 0.0001$\\
      \hline
      log(K) (m~s\textsuperscript{-1}) & $\mathcal{U}$(0.0, 4.0)& Table \ref{tab:planet}\\
    \textit{HARPS} offset & $\mathcal{N}$(7442.50, 10.0)& $7442.8 \pm 0.9$\\
    log(\textit{HARPS} jitter) & $\mathcal{N}$(1.895, 5.0)& $-2.27 ^{+2.37}_{-3.43}$\\
    \hline
    LD coefficient $u_1$ & \cite{exoplanet:kipping13}$^{\dag}$ & $0.34 \pm 0.30$\\
    LD coefficient $u_2$ & \cite{exoplanet:kipping13}$^{\dag}$ & $0.31 \pm 0.34$\\
      Orbital period (days) & $\mathcal{N}$(3.737882, 0.001) & Table \ref{tab:planet}\\
    T\textsubscript{C}\ (TBJD) & $\mathcal{N}$(2204.4183, 0.01) &Table \ref{tab:planet}\\
    log(R\textsubscript{p}) (R$_{\odot}$) & $\mathcal{N}$(-3.05, 1.0) & Table \ref{tab:planet}\\
    $[\sqrt{e}\,sin(\omega),\sqrt{e}\,cos(\omega)]$ & unitdisk$^{\ddag}$ & Table \ref{tab:planet}\\
    Impact parameter & $\mathcal{U}$(0.0, 1.0) & Table \ref{tab:planet}\\
    
      \hline
    \end{tabular}
    $^{*}$ Proportion of the measured flux from TESS that is due to TOI-2498.
      \\

      $^{\dag}$ The distributions for limb darkening coefficients are built into the \texttt{exoplanet} framework and are based on \cite{exoplanet:kipping13}.
      \\
      
      $^{\ddag}$ Eccentricity and the argument of periastron are sampled with a Uniform distribution on a unit disk in two dimensions using \texttt{pmx.unitdisk} from \texttt{PYMC3} \citep{exoplanet:pymc3}. 
      \\
      
      $\mathcal{N}(\mu, \sigma)$: Normal distribution with mean $\mu$ and standard deviation $\sigma$. 
      

      $\mathcal{U}(a,b)$: Uniform distribution with lower bound $a$ and upper bound $b$.


  \label{tab:priors}
\end{table}

\begin{table}
\renewcommand{\arraystretch}{1.75}

    \caption{TOI-2498\,b parameters}
    \begin{tabular}{l|l}
    \hline 
    {\bf Fitted Parameters} & {\bf Value}\\
    \hline

    Orbital period (days) & $3.738252 \pm 0.000004$\\
    Radius (R$_{\oplus}$) &  $6.06^{+0.29}_{-0.27}$\\
    R\textsubscript{p}/R\textsubscript{*} & $0.04 \pm 0.01$\\
    T\textsubscript{C}\ (TBJD) &  $2204.4167 \pm 0.0001$\\
    Impact parameter &  $0.42 \pm 0.13$\\
    \textit{K} (m~s\textsuperscript{-1}) &  $13.25 \pm 1.36$\\
    Eccentricity &  $0.089 \pm 0.075$\\
    Argument of periastron ($^{\circ}$) &  $89.99 \pm 99.14$\\
    \hline 
    {\bf Derived Parameters} & {\bf Value}\\
    \hline
    Mass (M$_{\oplus}$) & $34.62^{+4.10}_{-4.09}$\\
    Density (g cm\textsuperscript{-3}) &  $0.86^{+0.25}_{-0.20}$\\
    Inclination ($^{\circ}$) & $87.12 \pm 1.70$\\
    Semi-major axis (AU) &  $0.0491 \pm 0.003$\\
    Temperature T\textsubscript{eq}\ (K) $^*$  &  $1443 ^{+15}_{-28}$\\    
    \hline
    \end{tabular}
    
    $^{*}$ Assuming Albedo = 0 and uniform surface temperature

    \label{tab:planet}
    
\end{table}

\subsection{Transit timing variations}

We perform an initial check for additional planets using the DACE platform. Using DACE, we generate a lomb-scargle periodogram of the {\it HARPS} data and search for additional signals after removal of the radial velocity signal caused by TOI-2498\,b. We find no evidence of additional signals above a False Alarm Probability (FAP) of 0.1, suggesting that TOI-2498\,b is the only massive, short period planet orbiting TOI-2498. 

\begin{figure}
\centering

\includegraphics[width=\columnwidth, trim=0cm 0cm 0cm 0cm, clip]{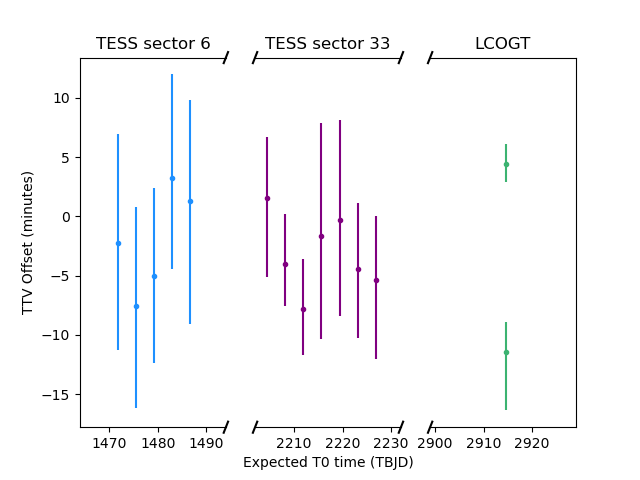}
\caption{TTV offsets for all transits from {\it TESS} sector 6 and sector 33, and LCOGT transits from CTIO and Teide Observatory.}
\label{fig:TTV}
\end{figure}

To support this analysis, we examine each individual full transit from {\it TESS} and LCO and check for transit timing variations. We re-run the transit model described in section \ref{subsection:photometry} for each individual transit. We fix all of the parameters to the best fit results in Table \ref{tab:planet}, other than the central transit time T0, for which we set a uniform prior, centered around the expected value. The results of this fitting are subtracted from the expected value to find the TTV offset for each transit. We present this analysis in Figure \ref{fig:TTV}. We again find no evidence of additional planets, although more transits would allow a more thorough check. It is worth noting the apparent discrepancy between the two LCOGT TTV offsets, as these points are different observations of the same transit and therefore should agree. We attribute this discrepancy to high airmass having a significant impact on the ingress of one of the transits (see Figure \ref{fig:LCO}). 

\section{Discussion}
\label{results}

\subsection{Planet position in the Neptune Desert}

\begin{figure}
\centering
\includegraphics[width=\columnwidth]{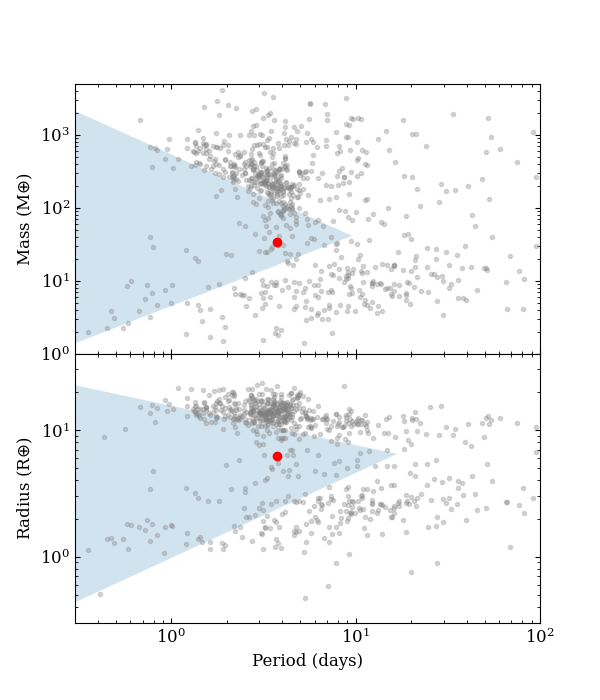}
\caption{The Neptunian desert shown in both mass-period (top) and radius-period (bottom) parameter space. The shaded blue regions indicate the areas within the boundaries laid out by \protect\cite{mazeh2016dearth}. The grey circles represent known exoplanets with mass uncertainties smaller than 30\% and radius uncertainties smaller that 10\%. The position of TOI-2498\,b is depicted by a larger red circle.} 
\label{fig:desert}

\end{figure}

We find that TOI-2498\,b is situated just within the Neptune desert, close to the lower edge in the mass-period parameter space. The positioning is displayed in Figure \ref{fig:desert} in both the mass-period and radius-period parameter spaces, with boundaries as given in \cite{mazeh2016dearth}.

As explained in \cite{owen2018photoevaporation}, it is thought that the upper boundary is due to high-eccentricity migration, while the lower boundary is due to photoevaporation. \cite{mazeh2016dearth} also provides an alternative explanation for the lower boundary, suggesting that as the planet-star separation grows, the size of the Hill sphere of the forming planet, the orbital path and the dust-to-gas ratio all increase. This results in an increased core mass at the end of the first stage of planetary formation. Hence planets that are further from their host stars have an increased total mass. 

As is apparent from Figure \ref{fig:desert}, the upper boundary of the Neptune desert is significantly more defined than the lower. There is an observational bias to take into account here, where larger, higher mass planets are substantially easier to detect and characterise. Consequently, as more planets close to the lower edge of the mass-period desert (such as TOI-2498\,b) are added to the sample, the boundary becomes even more blurred. This raises the question of whether the lower boundary of the Neptune desert needs to be re-evaluated. However, it is worth pointing out that programs such as HARPS-NOMADS specifically target planets within the desert parameter space, creating a biased sample of planets and potentially providing some explanation for this blurring of the boundary.

\subsection{Planet density}

\begin{figure*}
    \centering

    \includegraphics[width=\textwidth, trim=10cm 10cm 10cm 5cm, clip]{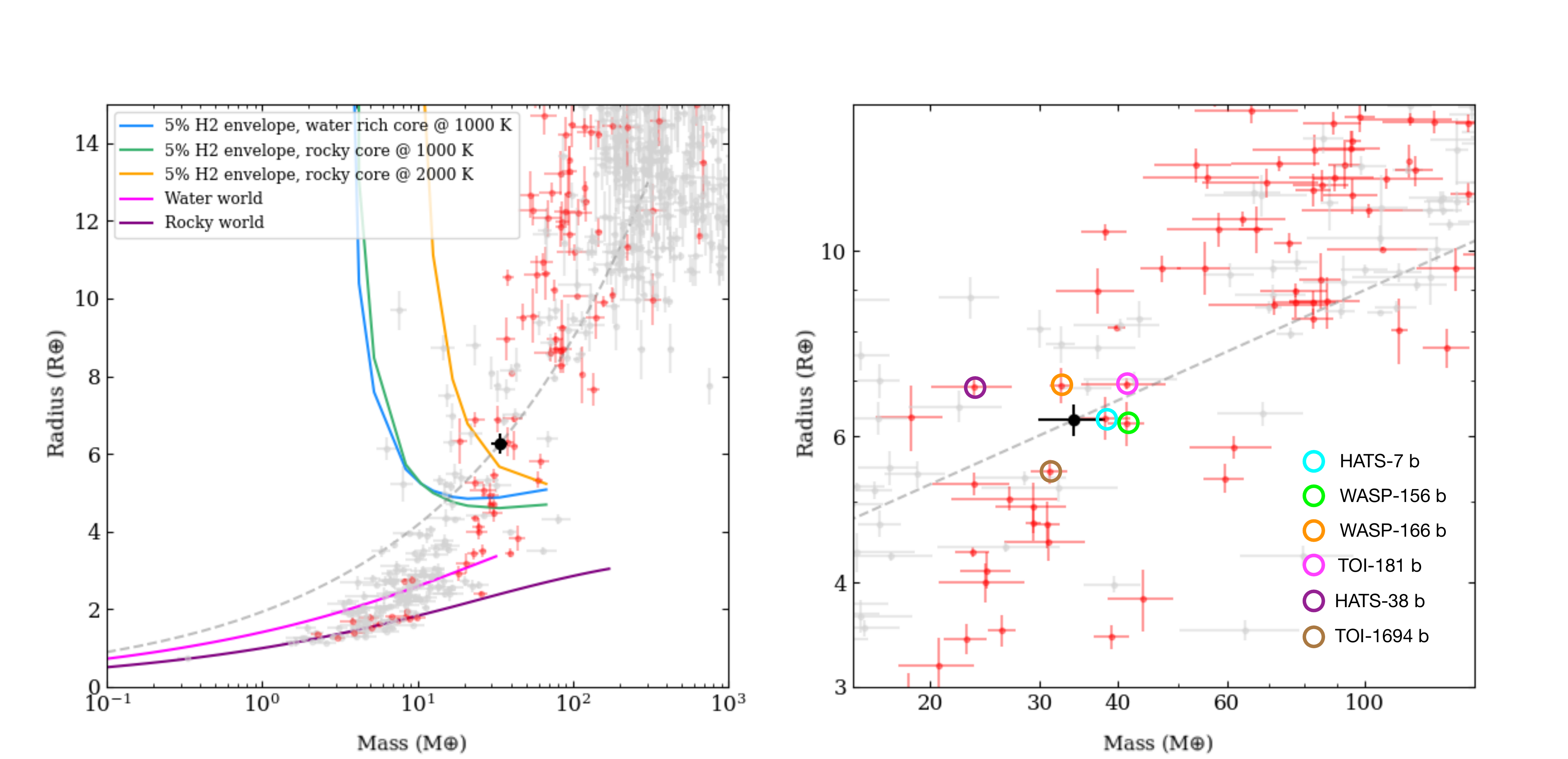}
    \caption{Position of TOI-2498\,b in mass-radius parameter space. In both plots, TOI-2498\,b is depicted as a black circle and the points in the background show the positions of all other confirmed planets with a radius uncertainty below 10$\%$ and a mass uncertainty below 25$\%$. The points highlighted in red are planets that are located within the Neptunian desert as defined by \protect\cite{mazeh2016dearth}. The grey dotted line is a line of constant density at 0.758 gcm\textsuperscript{-3} (the calculated density of TOI-2498). The left plot shows the positioning within a large parameter space and includes composition models from \protect\cite{zeng2016mass, zeng2019growth}. The right plot highlights planets within the neptune desert that are close to TOI-2498\,b within mass-radius parameter space. Note that the right plot is shown in log-log scale.}
    \label{fig:mass_radius}
\end{figure*}

We plot the position of TOI-2498\,b in mass-radius parameter space in Figure \ref{fig:mass_radius}. When considering the general exoplanet population, it can be seen that TOI-2498\,b sits in a relatively sparse area of this parameter space. This Figure also shows the planets proximity to composition models from \cite{zeng2016mass, zeng2019growth}, with the closest relation being the model for planets with a 5\% H$_{2}$ envelope and a rocky core with a surface equilibrium temperature of 2000~K. 

We also present the position of TOI-2498\,b in a more mass-constrained mass-radius space in the right-hand side of Figure \ref{fig:mass_radius}. In this plot, planets that are located within the Neptune desert are highlighted in red. Note that for these planets, there is an apparent upper size bound that falls rapidly as mass decreases. This could be a reflection of planets' on a closer orbit (shorter period) inability to maintain their envelopes due to irradiation from host stars.

We highlight 6 Neptune desert planets closest to TOI-2498\,b in mass-radius parameter space in Figure \ref{fig:mass_radius}. These planets are as follows: HATS-7 b \citep{HATS-7b}, WASP-156 b \citep{WASP-156b}, WASP-166 b \citep{WASP-166b}, TOI-181 b \citep{TOI-181b}, HATS-38 b \citep{HATS-38b}, TOI-1694 b \citep{TOI-1694b}. 

Of particular interest here is WASP-166 b, as not only does it share a very similar mass and radius to TOI-2498\,b, it also has a similar orbital period and equilibrium temperature. These similarities suggest that there is potential insight to be gained from a comparative study between atmospheres of TOI-2498\,b and WASP-166 b. Several atmospheric studies have been performed on WASP-166 b, in particular searches for sodium \citep{WASP-sodium} and water \citep{WASP-water}. We therefore suggest that TOI-2498\,b is a potential candidate for searching for species such as water or sodium in the optical using high-resolution, and also perhaps for carbon based molecules in the infrared using JWST.

\subsection{Composition and Internal structure}
\label{sec:internal-structure}

TOI-2498\,b's low density of $0.86$\,g\,cm$^{-3}$ is indicative of the presence of a gaseous envelope.
We estimate its internal structure by assuming it is comprised of a rocky core surrounded by an H/He-rich atmosphere, following \citet{RogersOwen21:radius-valley-origin}.
This model consists of four parameters: the rocky core radius R$_\text{core}$ and mass M$_\text{core}$, and the envelope thickness R$_\text{env}$ and mass fraction f$_\text{env}$. The envelope mass fraction is defined as the ratio of envelope mass to total planet mass $\text{f}_\text{env} = \text{M}_\text{env}/\text{M}_\text{p} = (\text{M}_\text{p} - \text{M}_\text{core})/\text{M}_\text{p}$. 

To solve for these parameters, we adopted a system of four equations.
These include (1) the definition of envelope mass fraction above, (2) the definition of envelope thickness as the difference between planet and core radii $\text{R}_\text{env} = \text{R}_\text{p} - \text{R}_\text{core}$, (3) the mass-radius relations for rocky cores by \citet{Otegi20:mass-radius}, and (4) the envelope structure model by \citet{ChenRogers16:envelope-model}, which determines the thickness of the gaseous envelope from its mass. The resulting internal structure of TOI-2498\,b, as shown in Table\,\ref{tab:internal-structure}, indicates that the planet hosts a significant gaseous envelope which consists of $27\pm4$\% of its mass. However, this model assumes no water content or significant atmospheric metallicity. Whilst massive planets with high water content are thought to have a mixed interior with a moderate metallicity gradient \citep{ref1}, smaller dry sub-Neptunes are thought to settle into a structure with distinct core and envelope \citep{ref2}. TOI-2498\,b is intermediate in size and would likely fall somewhere between these two scenarios. Future spectroscopic observations could constrain the metallicity of its atmosphere and its H2O content.


\begin{table}
    \caption{TOI-2498\,b internal structure assuming a rocky core surrounded by a H/He-rich envelope.}
    \begin{tabular}{l|l|l}
    \hline 
    {\bf Property} & {\bf Symbol (Units)} & {\bf Value} \\
    \hline
    Core radius            & R$_\text{core}$ (R$_\oplus$) & $2.62\pm0.14$ \\
    Core mass              & M$_\text{core}$ (M$_\oplus$) & $25.12\pm3.41$ \\
    Envelope thickness     & R$_\text{env}$ (R$_\oplus$)  & $3.43\pm0.36$ \\ 
    Envelope mass fraction & f$_\text{env}$               & $0.27\pm0.04$ \\
    \hline
    \end{tabular}
    \label{tab:internal-structure}
\end{table}

\subsection{Evaporation history}

Atmospheric evaporation is thought to have a significant influence in sculpting exoplanet populations, giving rise to the Neptune desert \citep{OwenLai18:neptune-desert-origin} as well as the period-radius valley \citep[e.g.][]{LopezFortney12:envelope-model, OwenWu13:radius-valley-evap, JinMordasini14:valley-model, OwenWu17:radius-valley-evap}.
The underlying mechanisms for evaporation, however, are still unclear.
One of the proposed mechanisms is photoevaporation, where X-ray and extreme ultraviolet photons (together, XUV) emitted by stars are readily absorbed by the upper atmospheres of close-in planets, heating up the gas and driving a hydrodynamic wind that escapes the planet's gravitational well \citep[e.g.][]{Lammer03:xray-evap, Erkaev07:energy-limited, Kislyakova13:xuv-mass-loss}.
Additional mechanisms have also been proposed, such as core-powered mass loss \citep{Ginzburg16:core-powered, GuptaSchlichting20:core-powered}, in which the energy for evaporation is provided by the core's internal heat.

\begin{figure*}
\centering
\includegraphics[width=\textwidth] {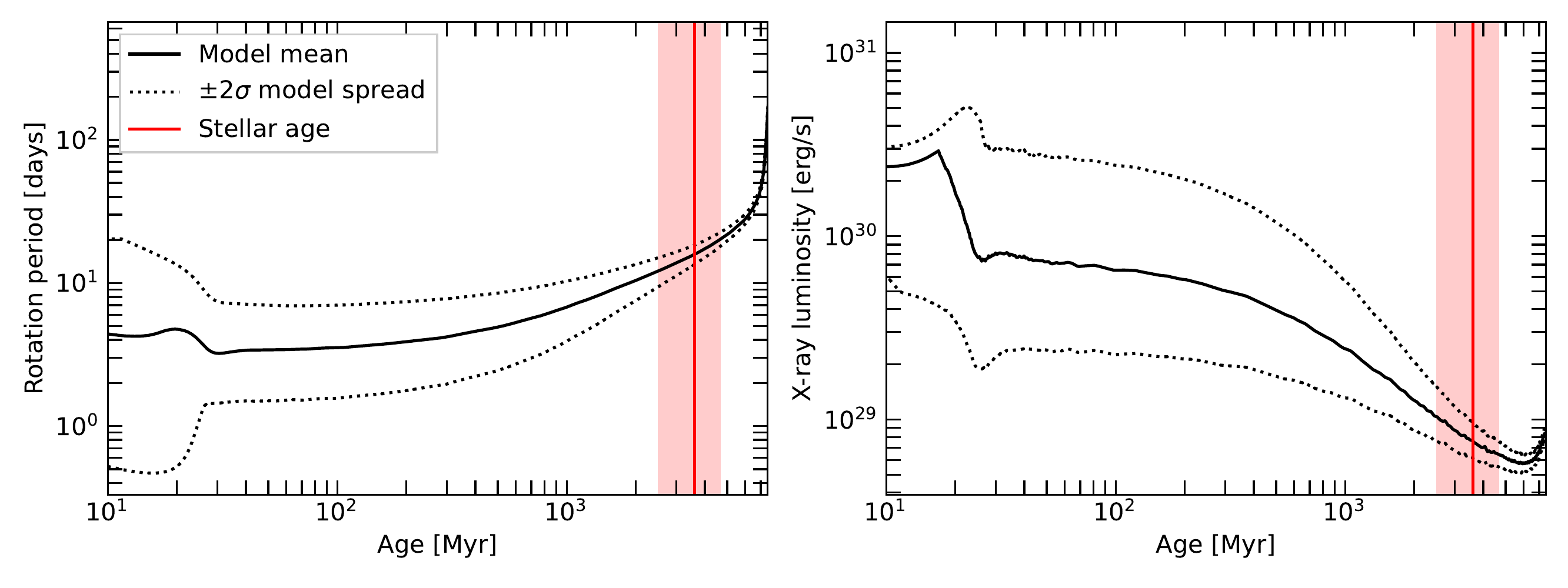}
\caption{{\bf Left panel}: spin evolution models for a 1.1\,M$_\odot$ star following the models by \citet{Johnstone21:rotation-model}. These include the mean and $2\sigma$ spread in starting rotation periods. The current age of TOI-2498, and its uncertainty, are shown as a red line and shaded region, respectively. {\bf Right panel}: models of the X-ray luminosity evolution for a 1.1\,M$_\odot$ star, akin to the left panel.}
\label{fig:star-evo}
\end{figure*}

Moreover, the XUV emission history of a star can be estimated from its spin period history, as the two quantities are related via the rotation-activity relation \citep{Wright11:rotation-xrays}, where a shorter stellar rotation period leads to a brighter X-ray emission.
We adopted the stellar spin evolution models by \citet{Johnstone21:rotation-model} to determine the X-ray emission past of TOI-2498.
For a given stellar mass, they modelled a distribution of starting rotation periods which is then evolved in time through various angular momentum transfer mechanisms.
The spin and X-ray evolution models for a 1.1\,M$_\odot$ star are shown on Figure\,\ref{fig:star-evo}. Even though we did not measure a rotation period for TOI-2498, we were able to determine an age of $3.6\pm1.1$\,Gyr, from which these models predict a present-day spin period between 8 and 24 days.

We modelled the evaporation history of TOI-2498\,b using the method by Fern\'{a}ndez Fern\'{a}ndez et al. (submitted).
We utilise the \texttt{photoevolver} code \footnote{The code is available on GitHub through: \url{https://github.com/jorgefz/photoevolver}} to simulate the evaporation history of the planet.
This method combines three ingredients: (1) a description of the stellar XUV emission history from the models by \citet{Johnstone21:rotation-model} as shown on Figure\,\ref{fig:star-evo}, (2) the evaporation model by \citet{Kubyshkina18:mass-loss-model}, which determines the mass loss rate from the incident XUV flux, and (3) the envelope structure model by \citet{ChenRogers16:envelope-model}, which recalculates the envelope thickness after mass is removed from it.
We evolved the planet under photoevaporation backwards in time to the age of 10 Myr and forwards to 7.5 Gyr.

\begin{figure*}
\centering
\includegraphics[width=\textwidth] {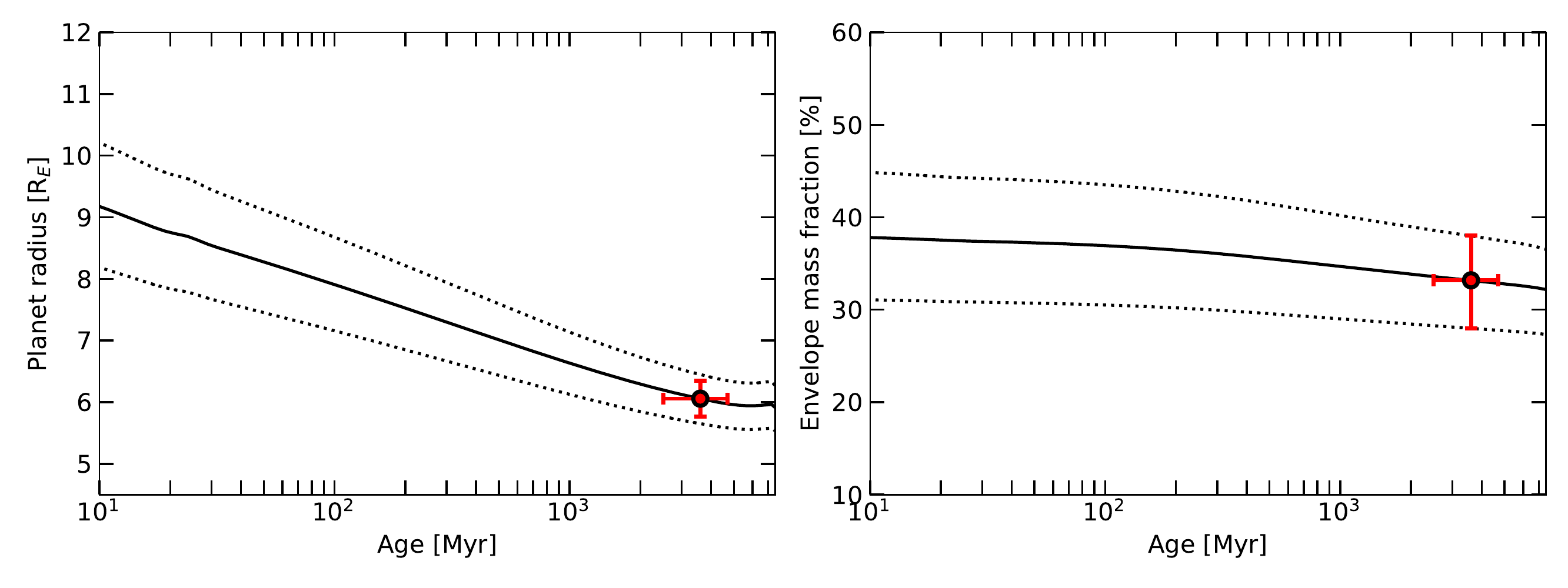}
\caption{{\bf Left panel}: plot of planet radius against age. The black lines represents the evolution of the radius of TOI-2498\,b under photoevaporation, both the mean (solid line) as well as $1\sigma$\,spread (dotted lines) based on the errors on the present-day mass, radius, and age, which are plotted as a red circle. {\bf Right panel}: plot of the planet's envelope mass fraction against age, following the left-hand panel.}
\label{fig:planet-evo}
\end{figure*}

Figure\,\ref{fig:planet-evo} shows the evaporation history of TOI-2498\,b.
Overall, we find TOI-2498\,b to be a stable super-Neptune that has experienced little evaporation throughout its lifetime, relative to its mass. The planet, after formation, might have started out as a puffy Saturn-sized world of radius 8--10\,R$_\oplus$ and envelope mass fraction 30--45\,\%. We also find its envelope is unlikely to be stripped prior to its host star evolving off the main sequence.


\section{Conclusions}
\label{conclusions}

In this work, we have presented the discovery of a hot, bloated Super-Neptune transiting a G type star. Our analysis includes photometry from {\it TESS} sector 6 and sector 33, follow up ground based photometry from LCOGT and spectroscopy from HARPS. We combine this data in a joint model to discern a planet radius of 6.1 ± 0.3 R\textsubscript{$\oplus$}, planet mass of 35 ± 4 M\textsubscript{$\oplus$} and an orbital period of 3.7 days. We calculate a stellar mass of 1.12 ± 0.02 M\textsubscript{$\odot$} and a stellar radius of 1.26 ± 0.04 R\textsubscript{$\odot$}. We find no indication of additional planets through radial velocity periodograms and TTV analyis. 

We estimate through internal structure models that 27$\pm$4\% of the mass of TOI-2498\,b constitutes a gaseous envelope. We present the evaporation history of the planet and find that it is likely a stable Super-Neptune, with little evaporation having taken place throughout its lifetime. We find that TOI-2498\,b likely started its life as a puffy Saturn-sized world before shrinking to its current size over 3.6 $\pm$ 1.1 Gyr.

We highlight similarities in mass, radius, period and temperature in TOI-2498\,b and WASP-166 b and suggest that there is potential for comparative atmospheric studies between the two planets. We also point out that spectroscopic atmospheric observations would allow for constraints on water content and atmospheric metallicity. This would improve our abilities to study the composition and internal structure of TOI-2498\,b.

\section*{Acknowledgements}

This work makes use of \texttt{tpfplotter} by J. Lillo-Box (publicly available in www.github.com/jlillo/tpfplotter), which also made use of the python packages \texttt{astropy}, \texttt{lightkurve}, \texttt{matplotlib} and \texttt{numpy}.

This research makes use of \texttt{exoplanet} \citep{exoplanet:exoplanet} and its dependencies \citep{exoplanet:agol20, exoplanet:arviz, exoplanet:astropy13,
exoplanet:astropy18, exoplanet:exoplanet, exoplanet:kipping13,
exoplanet:luger18, exoplanet:pymc3, exoplanet:theano}.

This work makes use of data from the European Space Agency (ESA) mission {\it Gaia} (\url{https://www.cosmos.esa.int/gaia}), processed by the {\it Gaia} Data Processing and Analysis Consortium (DPAC, \url{https://www.cosmos.esa.int/web/gaia/dpac/consortium}). Funding for the DPAC has been provided by national institutions, in particular the institutions participating in the {\it Gaia} Multilateral Agreement.

This paper includes data collected by the {\it TESS} mission. Funding for the {\it TESS} mission is provided by the NASA Explorer Program. Resources supporting this work were provided by the NASA High-End Computing (HEC) Program through the NASA Advanced Supercomputing (NAS) Division at Ames Research Center for the production of the SPOC data products. The {\it TESS} team shall assure that the masses of fifty (50) planets with radii less than 4 REarth are determined.

We acknowledge the use of public {\it TESS} Alert data from pipelines at the {\it TESS} Science Office and at the {\it TESS} Science Processing Operations Center.

This research makes use of the Exoplanet Follow-up Observation Program website, which is operated by the California Institute of Technology, under contract with the National Aeronautics and Space Administration under the Exoplanet Exploration Program.

This work makes use of observations from the LCOGT network. Part of the LCOGT telescope time was granted by NOIRLab through the Mid-Scale Innovations Program (MSIP). MSIP is funded by NSF.

This study is based on observations collected at the European Southern Observatory under ESO programme 1108.C-0697 (PI: Armstrong).

Based in part on observations obtained at the Southern Astrophysical Research (SOAR) telescope, which is a joint project of the Minist\'{e}rio da Ci\^{e}ncia, Tecnologia e Inova\c{c}\~{o}es (MCTI/LNA) do Brasil, the US National Science Foundation’s NOIRLab, the University of North Carolina at Chapel Hill (UNC), and Michigan State University (MSU).

We acknowledge the support from Funda\c{c}\~ao para a Ci\^encia e a Tecnologia (FCT) through national funds and from FEDER through COMPETE2020 by the following grants: UIDB/04434/2020 \& UIDP/04434/2020, 2022.04416.PTDC and 2022.06962.PTDC. E.D.M. acknowledges the support from FCT through Stimulus FCT contract 2021.01294.CEECIND. S.G.S. acknowledges the support from FCT through Investigador FCT contract nr.CEECIND/00826/2018 and POPH/FSE (EC). DJA is supported by UKRI through the STFC (ST/R00384X/1) and EPSRC (EP/X027562/1).

This work has been carried out within the framework of the National Centre of Competence in Research PlanetS supported by the Swiss National Science Foundation under grants 51NF40\_182901 and 51NF40\_205606. The authors acknowledge the financial support of the SNSF. This project has received funding from the European Research Council (ERC) under the European Union’s Horizon 2020 research and innovation program (grant agreement SCORE No. 851555) 

GF acknowledges a Warwick prize scholarship (PhD) made possible thanks to a generous philanthropic donation. GF and HMC acknowledge funding from a UKRI Future Leader Fellowship, grant number MR/S035214/1. FH is supported by an STFC studentship. PJW acknowledges support from the UK Science and Technology Facilities Council (STFC) through consolidated grant ST/T000406/1. KAC acknowledges support from the TESS mission via subaward s3449 from MIT.

\section*{Data Availability}

The {\it TESS} data can be accessed through the MAST (Mikulski Archive for Space Telescopes) portal at \url{https://mast.stsci.edu/portal/Mashup/Clients/Mast/Portal.html}. Photometry and imaging data from LCO and SOAR are accessible via the ExoFOP-TESS archive at \url{https://exofop.ipac.caltech.edu/tess/target.php?id=350153977}. All \texttt{python} scripts, including the \texttt{exoplanet} modelling code, parameter analysis and plotting can be made available upon reasonable request to the author.



\bibliographystyle{mnras}
\bibliography{bib} 



\bsp	
\label{lastpage}
\end{document}